# Tunable Metamaterial Absorber Based on Archimedean Spiral-Shaped Structure

Mahyar Radak[*], Saeed Mirzanejad

Department of Atomic and molecular physics, Faculty of Sciences, Mazandaran University, Babolsar, Iran
E-mail: M.Radak03@umail.umz.ac.ir *

## Abstract

In recent times, the Archimedean spiral structure has been considered as a promising design element in construction for specific purposes and opening up new possibilities in various applications. Its distinctive geometry exhibits a continuous growth pattern with a constant separation between its successive turns. One notable application of the Archimedean spiral structure in metamaterial absorbers is in achieving broadband absorption. This paper presents a comprehensive simulation of a tunable metamaterial absorber with an Archimedean spiral structure in the frequency range of 60 to 600 terahertz. The absorber's absorption spectrum is controlled by temperature variations that induce changes in the conductivity of vanadium dioxide. The absorber is composed of three layers: a bottom layer made of gold, a middle layer consisting of vanadium dioxide, and an upper layer is constructed using a gold cylinder, from which the Archimedean spiral with the same thickness as the gold is subtracted. This research provides valuable insights into the design and optimization of tunable metamaterial absorber.

**Keywords:** Archimedean spiral, metamaterial absorber, vanadium dioxide (VO2)

## Introduction

Currently, IR technology is commonly applied in a wide range of applications, including thermal imaging [1], sensing [2], radiation detectors [3], and energy harvesting [4]. The perfect metamaterial absorber was first proposed by Landy and his colleagues in 2008 [5] and in the following years, various structures for the perfect absorber such as the L-shaped dual-band [6], T-shaped double-band symmetric and full multi-absorbers with different structures [7] were proposed. These metamaterial absorbers are utilized in various fields such as stealth technologies [8], wireless communications [9], defense [10], and biomedical applications [11]. In the infrared spectrum, perfect metamaterial absorbers are crafted to efficiently absorb and manipulate electromagnetic radiation in the range of longer wavelengths. These absorbers aim to maximize absorption efficiency while maintaining spectral selectivity. These absorbers offer unique properties for controlling and manipulating electromagnetic radiation across different spectral ranges [12-16]. Thermal tunability relies on temperature-responsive materials [17-18], such as VO2 [19-20]. These materials undergo sudden changes in chemical structure (crystalline to amorphous) or material state (dielectric to metal) in response to temperature. They can be used as part of a layer or as an entire layer in metamaterial with additional processing. Thermal tunability is a method for adjusting material properties based on their response to changes in temperature. This technique relies on temperature-responsive materials such as VO2 [21] and germanium-antimony-telluride (GST) [22]. These materials undergo abrupt changes in their chemical or material states, transitioning from a crystalline to an amorphous structure or from a dielectric to a metallic state when subjected to temperature variations. They can be incorporated as layers within metamaterials or used as complete layers, often requiring additional processing steps. In addition to these specific examples, other temperature-responsive materials used in thermally tunable metamaterials include specialized polymers [23], distilled water [24], and liquid crystals [25]. We have recently proposed a double tunable metamaterial based on vanadium oxide and PZT (Lead Zirconate Titanate), which enables absorption by simultaneously changing two parameters [26]. In this article, we introduce a novel tunable metamaterial absorber that harnesses the unique geometry of the Archimedean spiral structure. This absorber offers tunability in absorption properties, enabling precise control over the absorption spectrum for various applications. By carefully engineering the dimensions and parameters of the spiral, researchers have been able to design absorbers with enhanced absorption efficiency over a wide bandwidth. The Archimedean spiral, defined by the polar equation $r = a + b*\theta$, has proven to be a remarkable design with diverse applications. Its unique geometry enables broadband absorption, making it well-suited for situations that require the absorption of a wide range of wavelengths. This characteristic has led to its utilization in various fields, including antenna design [27-30], wind turbines [31-32], acoustic vortex generation [33-34], wireless sensor system [35] and the fabrication of metamaterials with negative refractive index [36]. Also introduced recently was the concept of the optical Archimedes screw, a new system capable of capturing and amplifying light inspired by the traditional Archimedes screw mechanism [37]. In recent times, researchers have proposed metamaterial absorbers based on the Archimedean structure, thanks to its intriguing properties [38-39]. In this article, we introduce a tunable metamaterial absorber that incorporates the Archimedean structure. By manipulating the parameters of the Archimedean spiral, we aim to investigate how these changes affect the absorption spectrum. Additionally, we will explore the impact of conductivity changes in vanadium dioxide on the absorption characteristics of the metamaterial. By adjusting the parameters of the spiral structure, such as the initial radius and tightness or looseness, we can tailor the absorber's performance to specific wavelength ranges. This tunability is crucial for applications that require absorption across a wide range of wavelengths. Additionally, we will investigate the absorption spectrum extent at various angles of incidence θ, as well as different azimuthal angles denoted by φ. By varying the angle of incidence and azimuthal, we can analyze how the absorber's performance varies for different incoming wave directions. This examination will provide insights into the absorber's angular response and its ability to absorb electromagnetic waves effectively from a range of incident angles

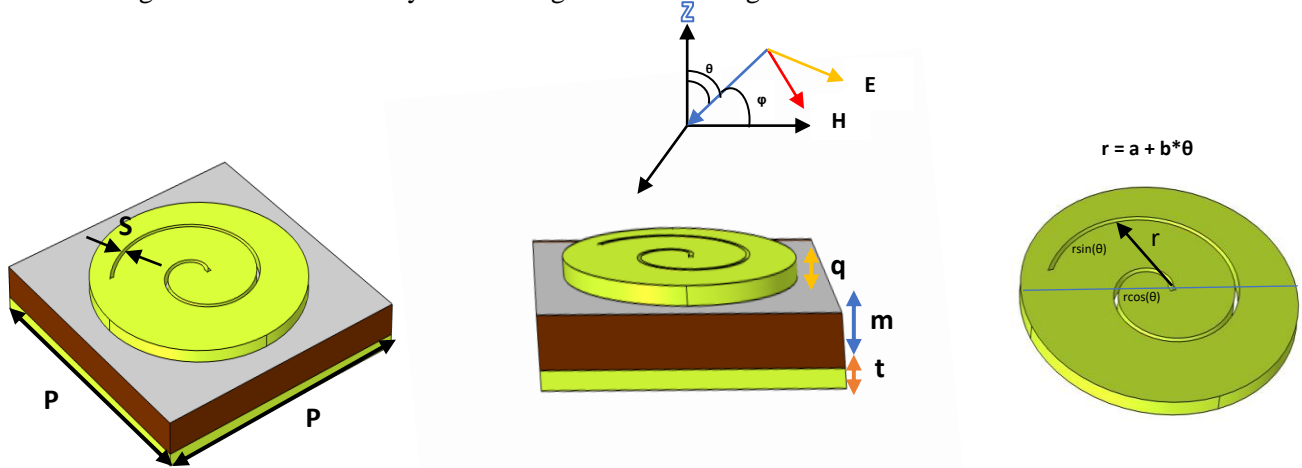

**Fig 1.** The proposed metamaterial absorbent structure has a size (p) of 2.5 μm. This design consists of three layers, the metal layers of gold with thicknesses of t = 0.1 μm and q = 0.1 μm and the middle layer of vanadium oxide with a thickness of m = 0.27 μm and the size of the Archimedes spiral s = 0.04 μm.

Figure 1 illustrates the structure of the metamaterial absorber, showcasing the arrangement of the layers and their respective dimensions. This unit cell design serves as the fundamental building block for the complete absorber structure and can be replicated to form an array, thereby enhancing the overall performance.

**Theoretical analysis and simulation**

When the temperature undergoes changes, vanadium dioxide (VO2), an active material, exhibits a phase transition temperature at 68 degrees. At this critical temperature, VO2 undergoes a transition from a dielectric to a conductor, resulting in an increase in its electrical conductivity. The optical properties of VO2 in the terahertz (THz) range are described using the Drude model, as expressed in Equation (1):

$$\varepsilon(\omega) = \varepsilon_\infty - \frac{\omega_p^2}{\omega(\omega+i\omega_c)} \tag{1}$$

$\varepsilon_\infty = 12$ represents the dielectric permittivity at high frequency, and $\omega_c = 5.75 \times 10^{13}$ rad/s is the collision frequency. The relationship between the plasma frequency ($\omega_p$) and conductivity ($\sigma$) can be described by Equation:

$$\omega_p^2(\sigma) = \frac{\sigma}{\sigma_0}\omega_p^2(\sigma_0) \tag{2}$$

with $\sigma_0 = 3 \times 10^5$ S/m and $\omega_p(\sigma_0) = 1.4 \times 10^{15}$ rad/s.

It is assumed that the conductivity of $VO_2$ changes from 200 S/m to $2 \times 10^5$ S/m when it undergoes a transition from the insulating phase to the metallic phase [29, 30]. In order to accurately characterize the optical properties of gold in this specific design, the Drude model is employed. The plasma frequency ($\omega_p$) for gold is determined to be $1.2 \times 10^{16}$ rad/s, while the collision frequency ($\omega_c$) is measured to be $10.5 \times 10^{13}$ rad/s. The absorption coefficient ($A(\omega)$) and reflection coefficient ($R(\omega)$) can be calculated using Equation:

$$A(\omega) = 1 - R(\omega) = 1 - |S11(\omega)|^2 - |S21(\omega)|^2 \tag{3}$$

where S11 and S21 represent the reflection parameter and transmission parameter, respectively [35]. Metamaterials, acting as effective mediums, can be characterized by complex electrical permittivity $\varepsilon(\omega) = \varepsilon_1(\omega) + i\varepsilon_2(\omega)$ and magnetic permeability $\mu(\omega) = \mu_1(\omega) + i\mu_2(\omega)$. The phenomenon of complete absorption in the absorber can be explained using impedance matching theory. The effective impedance of the perfect metamaterial absorber (PMA) can be derived from the effective permeability ($\mu$) and effective permittivity ($\varepsilon$) as described in Equation (4):

$$z = \sqrt{\mu/\varepsilon} = \sqrt{\frac{(1+S_{11}^2)-S_{21}^2}{(1-S_{11}^2)-S_{21}^2}} \tag{4}$$

where $\mu$ and $\varepsilon$ represent the effective permeability and permittivity, respectively [29].

The Archimedean spiral can be defined using polar coordinates (r, θ), where r represents the distance from the fixed point, and θ represents the angle of rotation. The equation that describes the spiral is:

$$r = a + b\theta$$

$$\begin{cases} x = (r + at)\cos(t) \\ y = (r + bt)\sin(t) \end{cases} \quad t \in (0, n\pi) \tag{5}$$

In this equation, b is a real number that determines the distance between successive loops of the spiral. By changing the value of b, you can control the spacing between the loops. As time elapses, the position of the particle from the starting point is proportional to the angle θ.

**Numerical results**

*A ) Tunability based on VO2 conductivity change*

In this section, we are exploring the relationship between the electrical conductivity of the vanadium dioxide (VO2) layer and the control it exerts over the absorption spectrum. Specifically, we are examining the effects of changing the electrical conductivity within a range of $2\times10^2$ s/m to $2\times10^5$ s/m. By systematically varying the electrical conductivity and observing the corresponding changes in the absorption spectrum, we can analyze the relationship between the two variables.

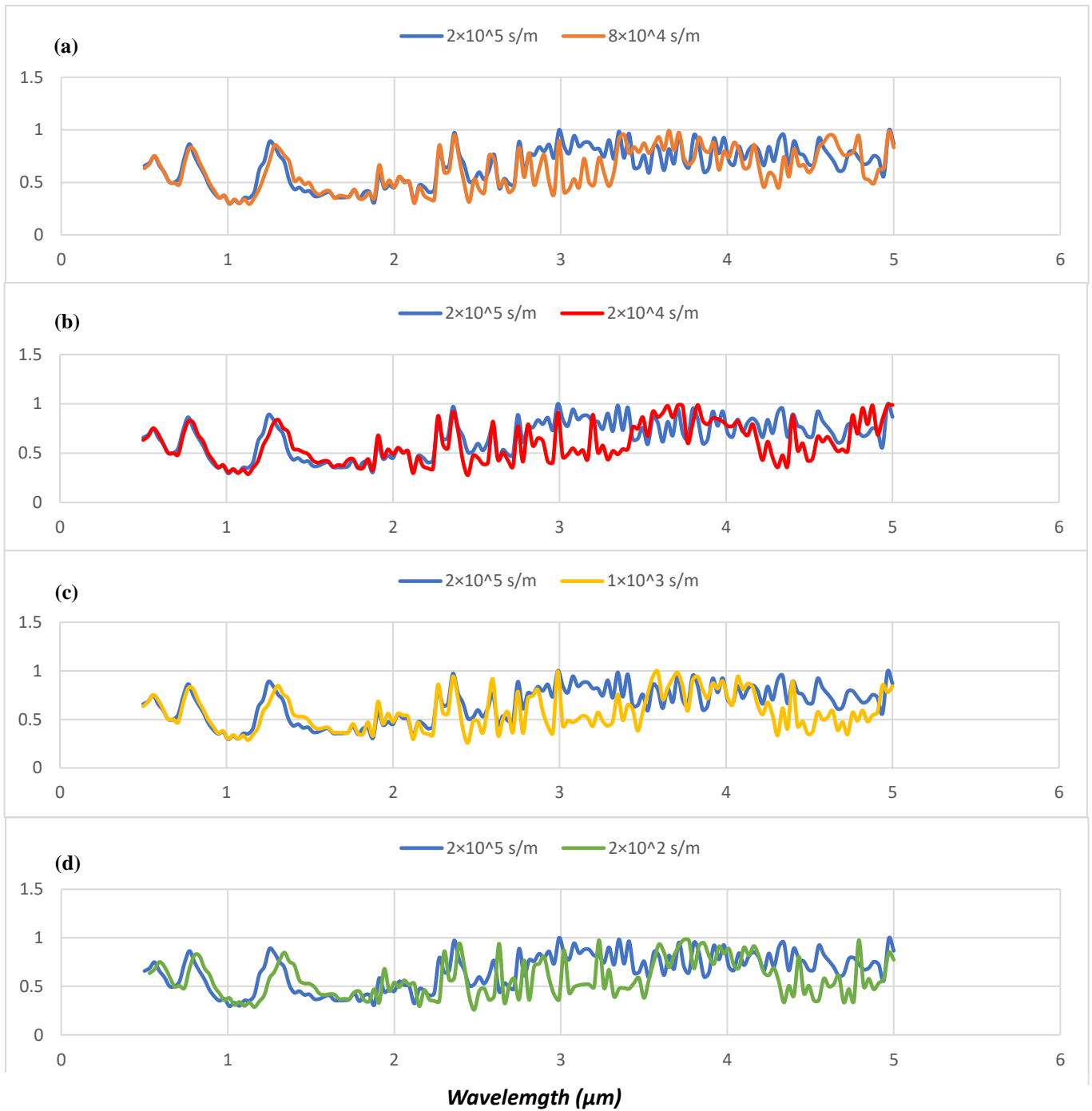

Wavelemgth (μm)

**Fig 2.** Analysis of the TE wave absorption spectrum of vanadium oxide (VO2) under different conductivity conditions, which is compared with the absorption spectrum diagram in conductivity $2\times10^5$ s/m, (a) $8\times10^4$ s/m, (b) $2\times10^4$ s/m, (c) $1\times10^3$ s/m, (d) $2\times10^2$ s/m.

In Figure 2, we observe that modulation of the electrical conductivity in vanadium dioxide leads to variations in the plasma frequency. Consequently, the absorption spectrum of the metamaterial absorber undergoes changes.

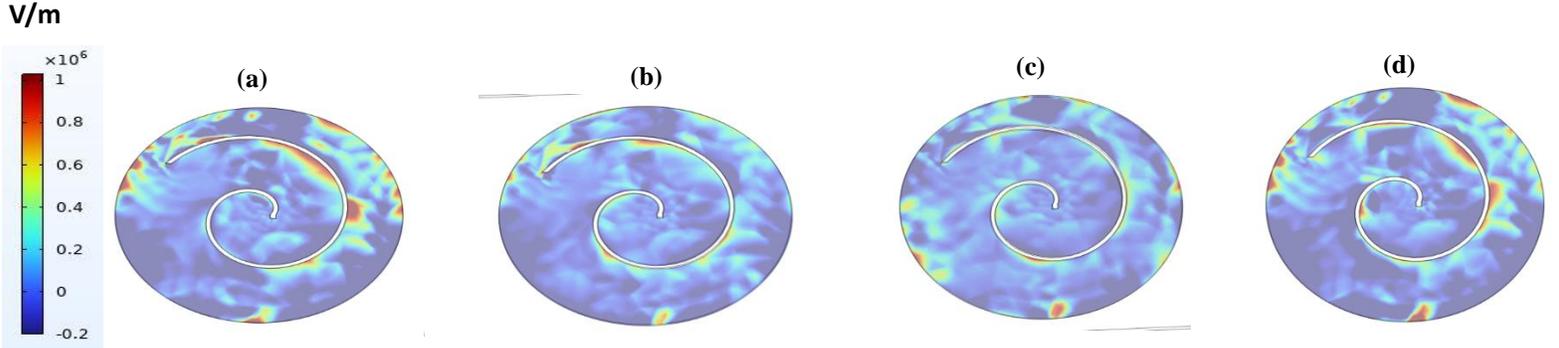

**Fig 3.** Z-component of electric field distribution on gold Archimedean spiral structure at λ = 2.99 μm, TE polarization, θ = 45° radiation angle, and φ = 0° for conductivity (a) $8\times10^4$ s/m, (b) $2\times10^4$ s/m, (c) $1\times10^3$ s/m, (d) $2\times10^2$ s/m.

The figure3 illustrates the z-component of the electric field distribution on the Archimedean spiral structure of a gold layer. The analysis focuses on the resonance wavelength λ = 2.99 μm and considers the TE (Transverse Electric) polarization mode. The radiation angle is set at 45 degrees, while the azimuthal angle φ is set to 0 degrees. The z-component of the electric field refers to the component of the electric field vector that is parallel to the z-axis of the coordinate system. In this case, it represents the electric field strength in the direction perpendicular to the gold layer. By examining the electric field distribution at the resonance wavelength and specific angles of radiation and azimuth, we gain insights into the behavior of the Archimedean spiral structure in response to TE waves.

### B) *The effect of Archimedes spiral structure on the absorption spectrum*

In this section, we aim to demonstrate the impact of the Archimedes spiral structure on enhancing the absorption spectrum for both TE and TM polarizations.

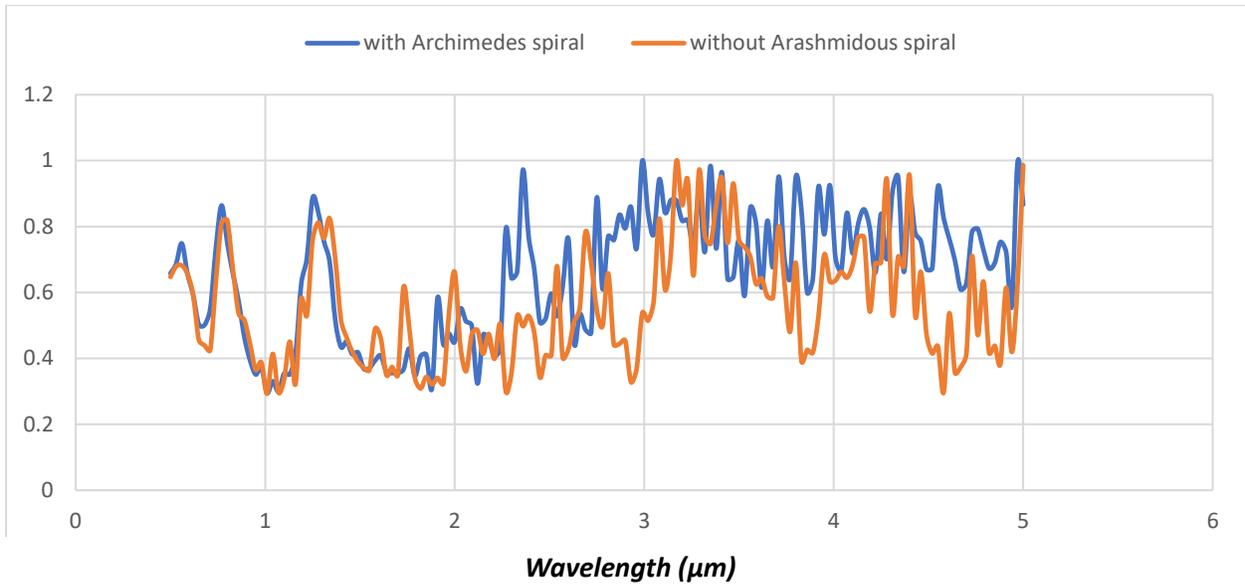

**Fig 4.** Comparison diagram of TE polarization absorption for metamaterial absorbent structure with and without Archimedean structure

In Figure 4, the results clearly illustrate notable changes in absorption levels. For TE polarization, the absorption peak experienced significant enhancement at specific wavelengths. At 2.36 micrometers, the absorption peak increased from 52% to 97%, demonstrating a substantial improvement. Similarly, at 2.99 micrometers, the absorption spectrum rose from 52% to 99%, indicating a remarkable increase in absorption efficiency. Furthermore, at 3.71 and 3.8 micrometers, the absorption rates surpassed 95%, signifying substantial improvements compared to the initial values of 79% and 68%, respectively. These findings demonstrate the pronounced effect of the Archimedean structure on enhancing absorption capabilities. Additionally, at a wavelength of 4.61 micrometers, the absorption showed a significant increase, rising from 53% to 92%. This further supports the overall conclusion that the presence of the Archimedean structure has a profound effect on the absorption spectrum of the absorptive structure. The findings presented in Figure 4 provide compelling evidence of the positive impact of incorporating the Archimedean structure on the absorption spectrum for TE polarization.

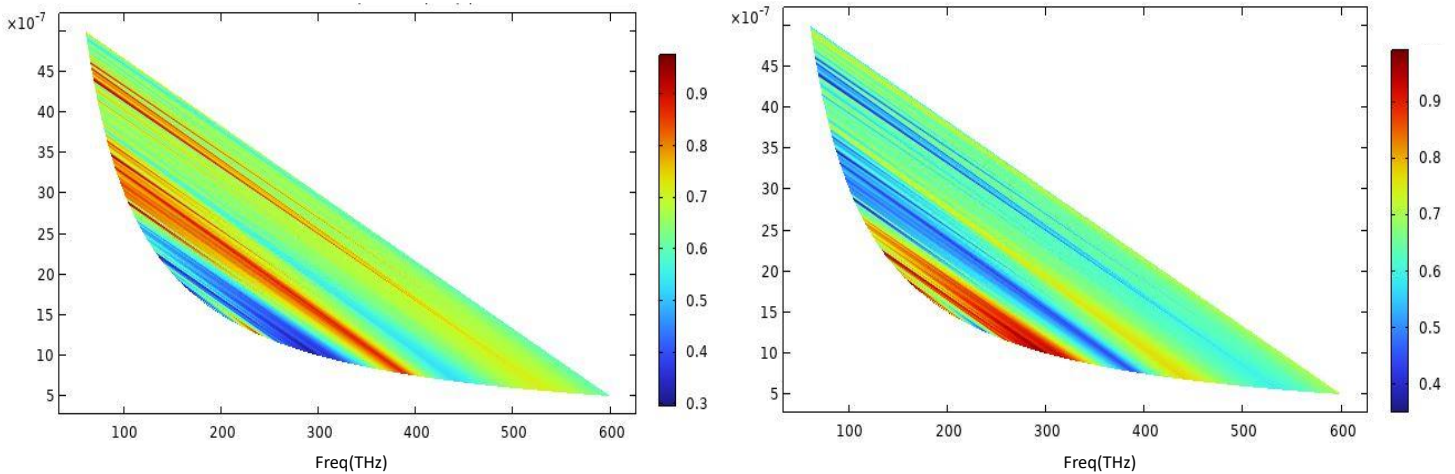

**Fig 5.** Absorption and Reflection Plot for TE Polarization

Figure 5 depicts the Absorption and Reflection Plot specifically for TE polarization. This plot provides a visual representation of the absorption and reflection coefficients as a function of the incident wave wavelength and frequency.

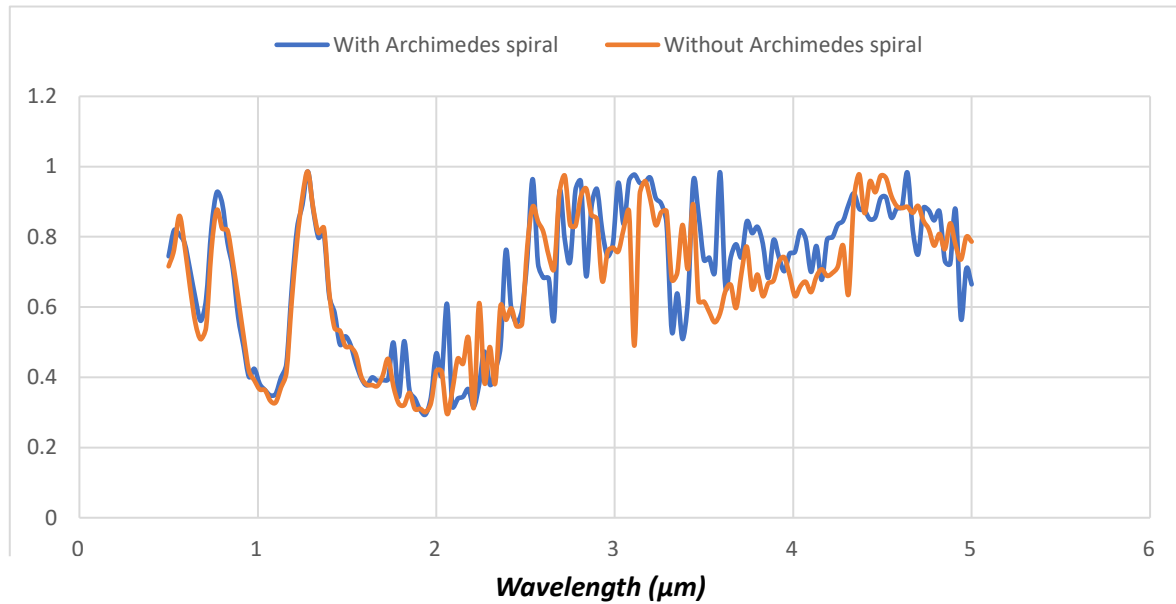

**Fig 6.** Comparison diagram of TM polarization absorption for metamaterial absorbent structure with and without Archimedean structure

In Figure 6, we present further investigations into the impact of the Archimedean structure on the polarization absorption spectrum of TM metamaterials. The application of the Archimedean structure resulted in notable enhancements in the absorption spectrum. For instance, at a wavelength of 2.54 μm, the absorption increased from 85% to 96%. Similarly, at 3.44 μm, there was an improvement from 89% to 95%. Notably, at 3.59 μm, a significant change was observed, with absorption rising from 59% to 98%. Furthermore, at 4.64 μm, absorption increased from 88% to 98%. The results highlight the substantial impact of the Archimedean structure on the polarization absorption spectrum of TE and TM metamaterials.

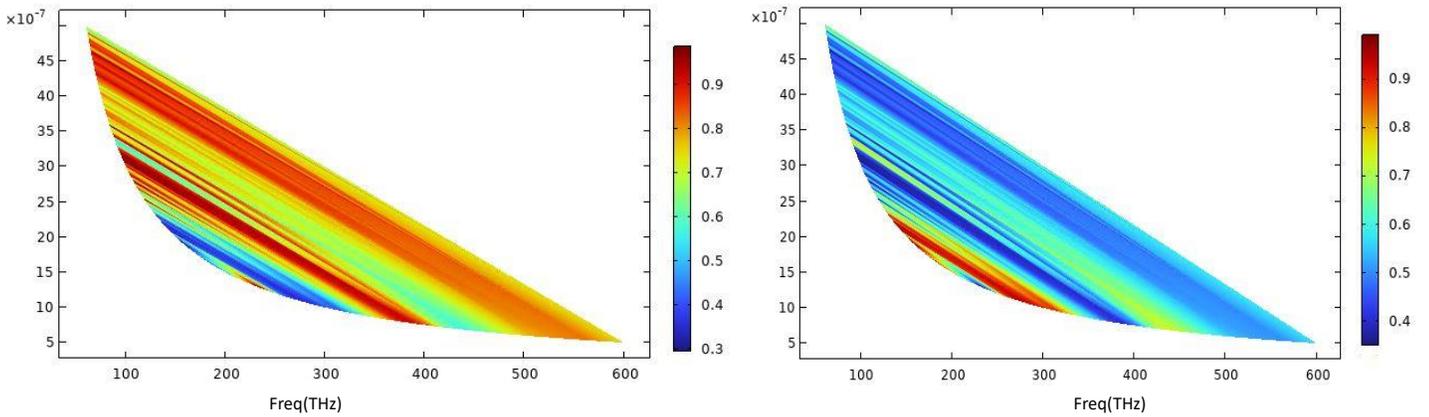

**Fig 7.** Absorption and reflection plot for TM polarization

Figure 7 depicts the Absorption and Reflection Plot specifically for TM polarization. This plot provides a visual representation of the absorption and reflection coefficients as a function of the incident wave wavelength and frequency.

### D) Dependence of absorption to the angle of incidence and azimuthal angle

In this section, we will explore the impact of absorption on both the incidence angle and the azimuthal angle within the range of 0 to 45 degrees.

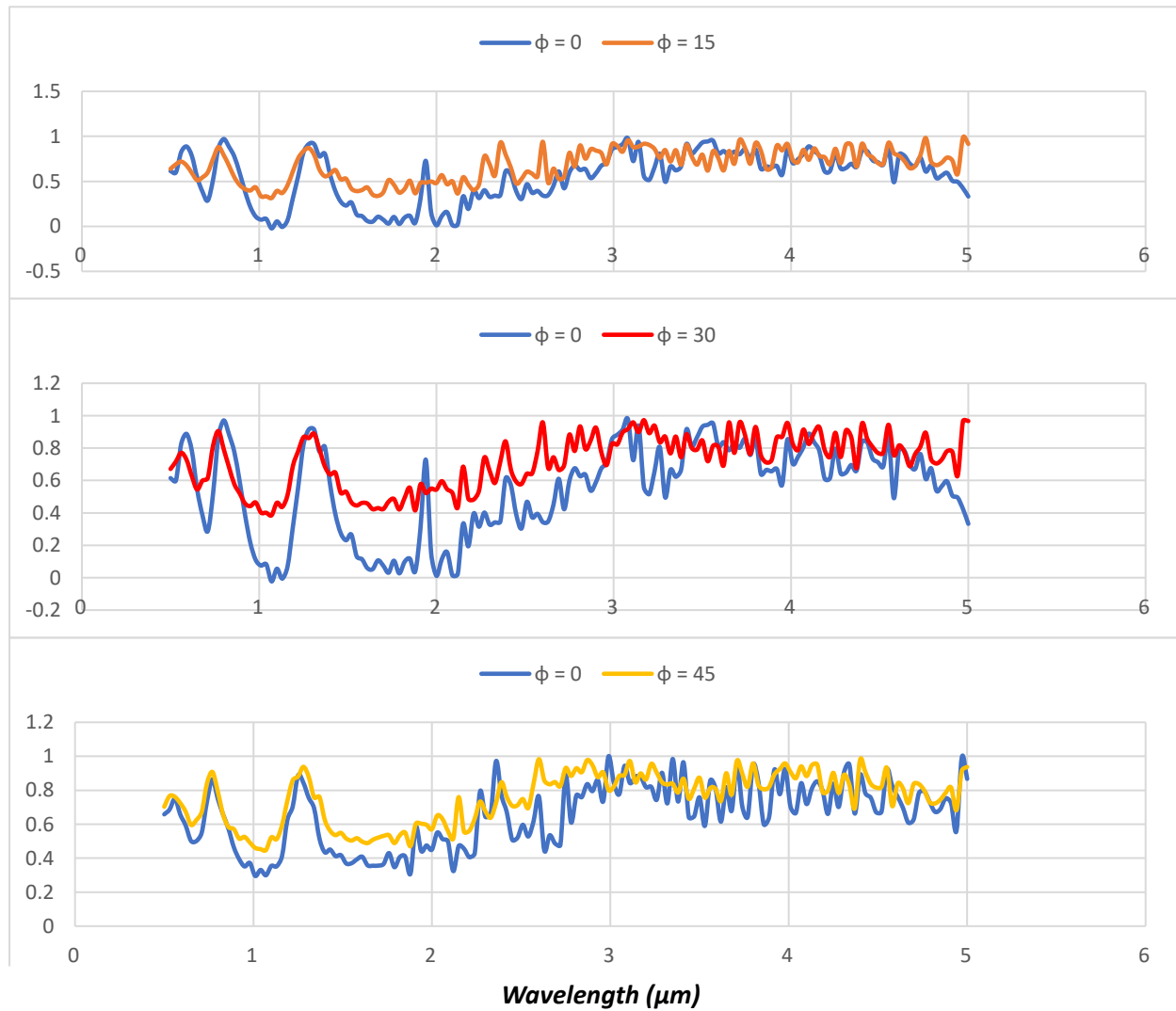

*Wavelength (μm)*

**Fig 8.** Absorption spectra of metamaterials at different azimuthal angle φ for TE configuration (θ = 45°, conductivity of vanadium oxide = 2×10$^5$ s/m)

In Figure 8, the experimental results demonstrate an intriguing characteristic of the absorber. The range of variation for the angle φ spans from 0 degrees to 45 degrees, while the angle θ remains fixed at 45 degrees. What makes this observation particularly interesting is the consistent and remarkable performance of the absorber across this angular span. Regardless of the angle at which the waves approach the absorber or their different orientations within this range, the absorber consistently maintains a high absorption efficiency. This means that whether the incident waves arrive at the absorber from a shallow angle or a steeper angle, or if they are oriented differently with respect to the absorber's surface, the absorber remains effective in capturing and absorbing the incident energy. This reliable and consistent absorption efficiency is of great significance in practical applications. It ensures that the absorber's performance is not affected by variations in the incidence angle or the orientation of the waves. This characteristic makes the absorber highly versatile, adaptable, and reliable in scenarios where the incidence waves may approach from different angles or orientations.

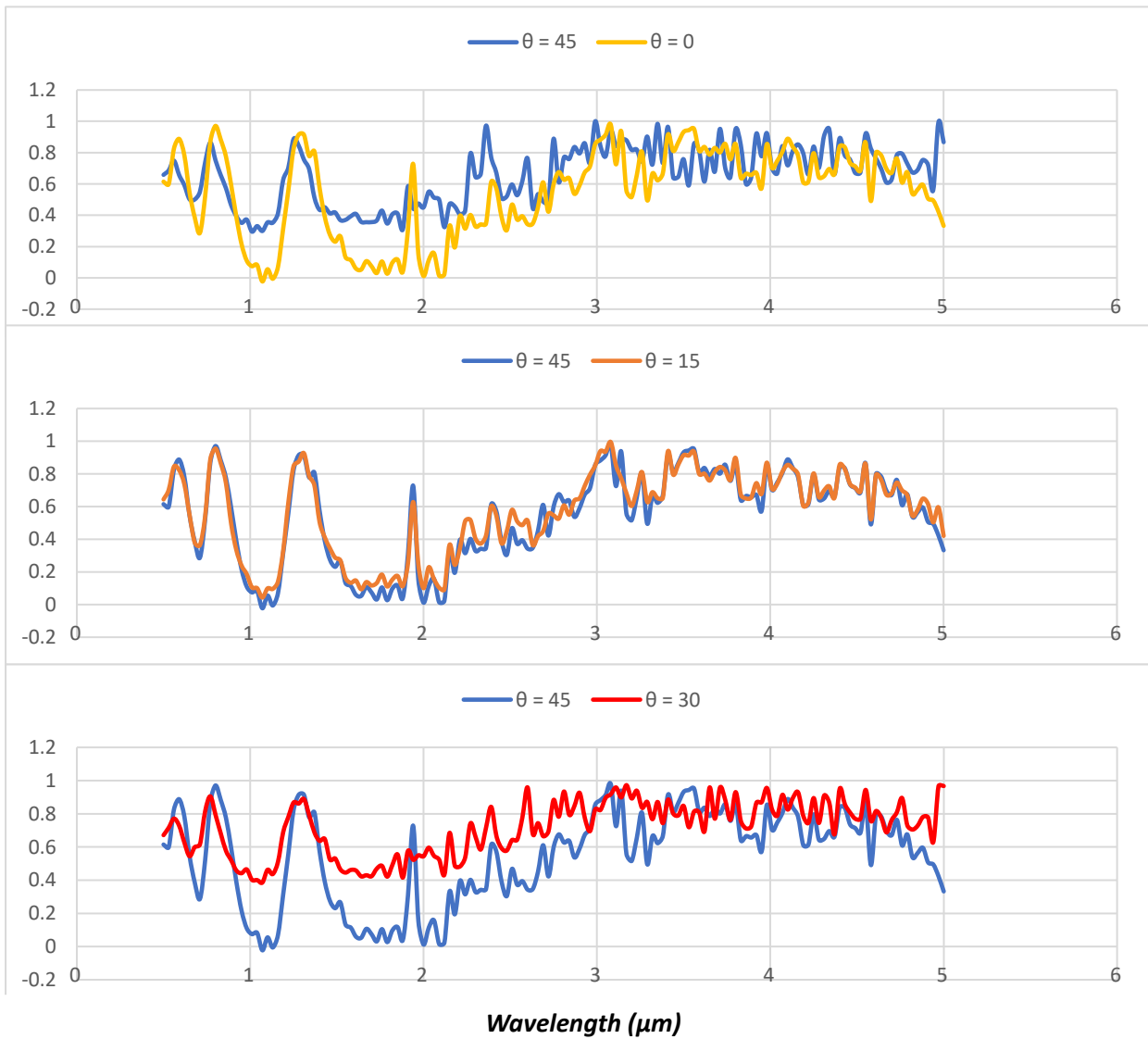

*Wavelength (μm)*

**Fig 9.** Absorption spectra of metamaterials at different azimuthal angle θ for TE configuration (φ = 0º, conductivity of vanadium oxide = 2×10$^5$ s/m)

In Figure 8, the influence of changing the azimuthal angle on the absorber's absorption spectrum is depicted. By altering the azimuthal angle, we observe notable variations in the absorption characteristics. Importantly, within the range of 0 to 45 degrees of azimuthal angle, the absorber consistently demonstrates good absorption performance. Figure 9, on the other hand, illustrates the impact of changing the theta angle on the absorption spectrum. Similar to the azimuthal angle variations, altering the θ angle leads to changes in the absorption performance of the absorber. Notably, within the range of 0 to 45 degrees of theta angle, the absorber exhibits favorable absorption characteristics. This implies that irrespective of the specific theta angle within this range, the absorber maintains good absorption efficiency, effectively absorbing incident waves. It is important to mention that during these variations in the θ angle, the angle φ is maintained at a constant value of 0 degrees. The results depicted in Figures 8 and 9 highlight the sensitivity of the absorber's absorption spectrum to changes in the azimuthal and incident angles, respectively. Within the specified ranges, both variations demonstrate the absorber's capability to maintain effective absorption performance.

### *d) Dependence of TE absorption wave on Archimedes spiral parameters*

Furthermore, we will explore the impact of varying the size of the Archimedean screw structure parameters on the absorption spectrum. By modifying parameters such as a, b, θ, and thickness S of the Archimedean structure, we aim to understand how these changes affect the absorber's absorption characteristics. By systematically varying the size of the Archimedean screw structure, we can observe how changes in these dimensions influence the absorber's ability to capture and absorb incident waves. This analysis will provide insights into the relationship between the structural parameters and the absorber's performance, enabling us to optimize the design for specific absorption requirements.

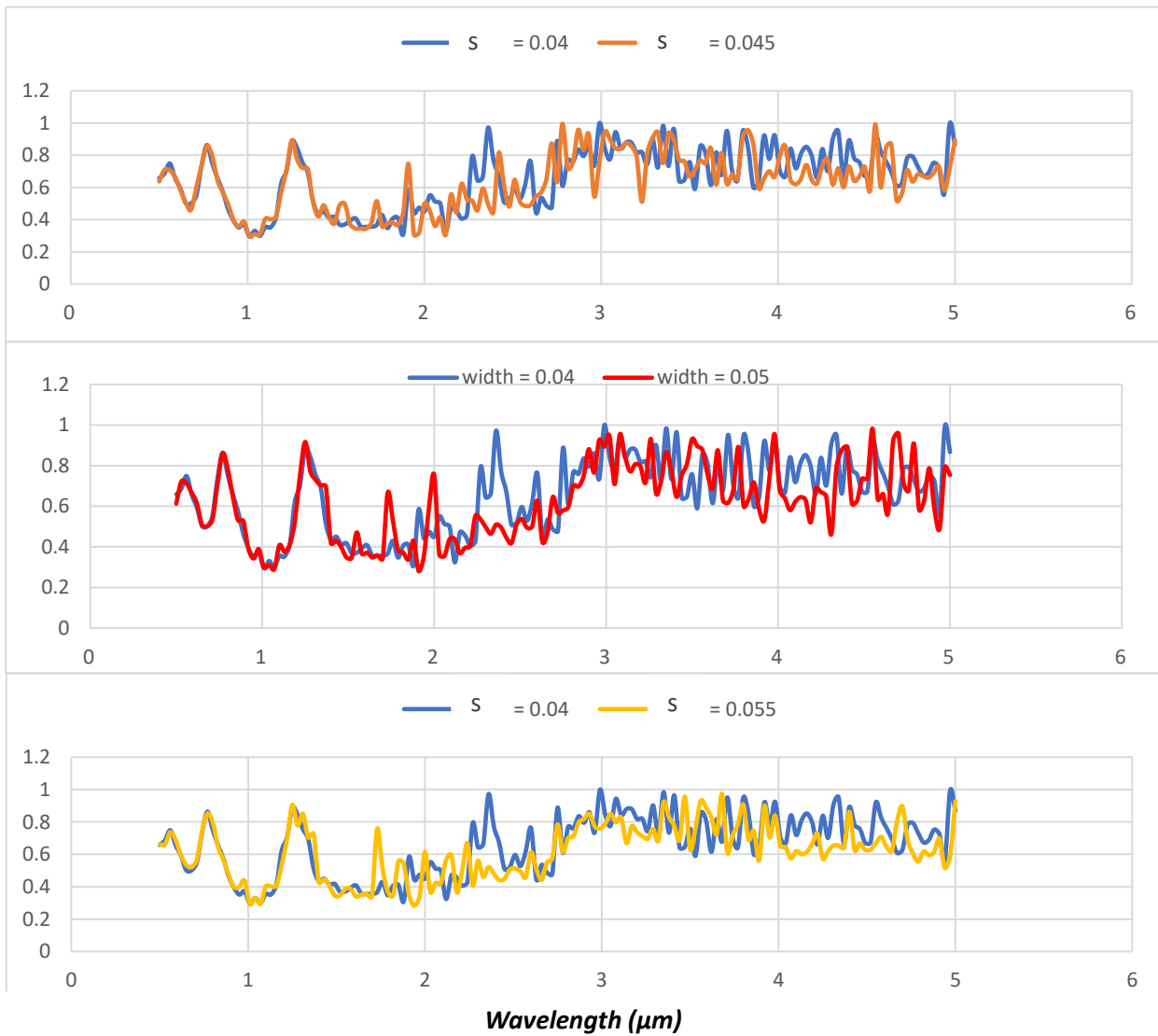

**Fig 10.** Investigating the effect of Archimedes screw thickness on absorption spectrum of metamaterial absorber

In Figure 10, we focus on investigating the effect of varying the thickness parameter (S) of the Archimedean screw on the absorption spectrum of the metamaterial absorber. By systematically changing the thickness of the Archimedean structure, we aim to understand how this parameter influences the absorber's absorption characteristics. Initially, the absorber is set with a thickness of S = 0.04 micrometers. By increasing the thickness to S = 0.045 micrometers, S = 0.05 micrometers, and S = 0.055 micrometers, we observe corresponding changes in the absorption spectrum. As the thickness of the Archimedean screw increases from its initial value of 0.04 micrometers, we analyze the resulting alterations in the absorption spectrum. As the thickness parameter (S) of the Archimedean screw increases, we observe a decrease in the absorption spectrum within our frequency range. Specifically, we note that the most optimal thickness is 0.04 micrometers, which corresponds to a remarkable absorption peak of 97% at a wavelength of 2.36 micrometers. When the thickness is increased to S = 0.045 micrometers, the absorption percentage decreases to 49%. Similarly, for a thickness of S = 0.05 micrometers, the absorption drops further to 50%, and for S = 0.055 micrometers, the absorption decreases to 52%. These results demonstrate a consistent trend of decreasing absorption efficiency as the thickness of the Archimedean screw increases. The correlation between thickness variations and the absorption spectrum provides valuable insights into optimizing the absorber's design. It indicates that the absorber's performance is most optimal when the thickness is set to 0.04 micrometers. Deviating from this optimal thickness leads to a decrease in absorption efficiency, as observed in the decreasing absorption percentages for higher thickness values.

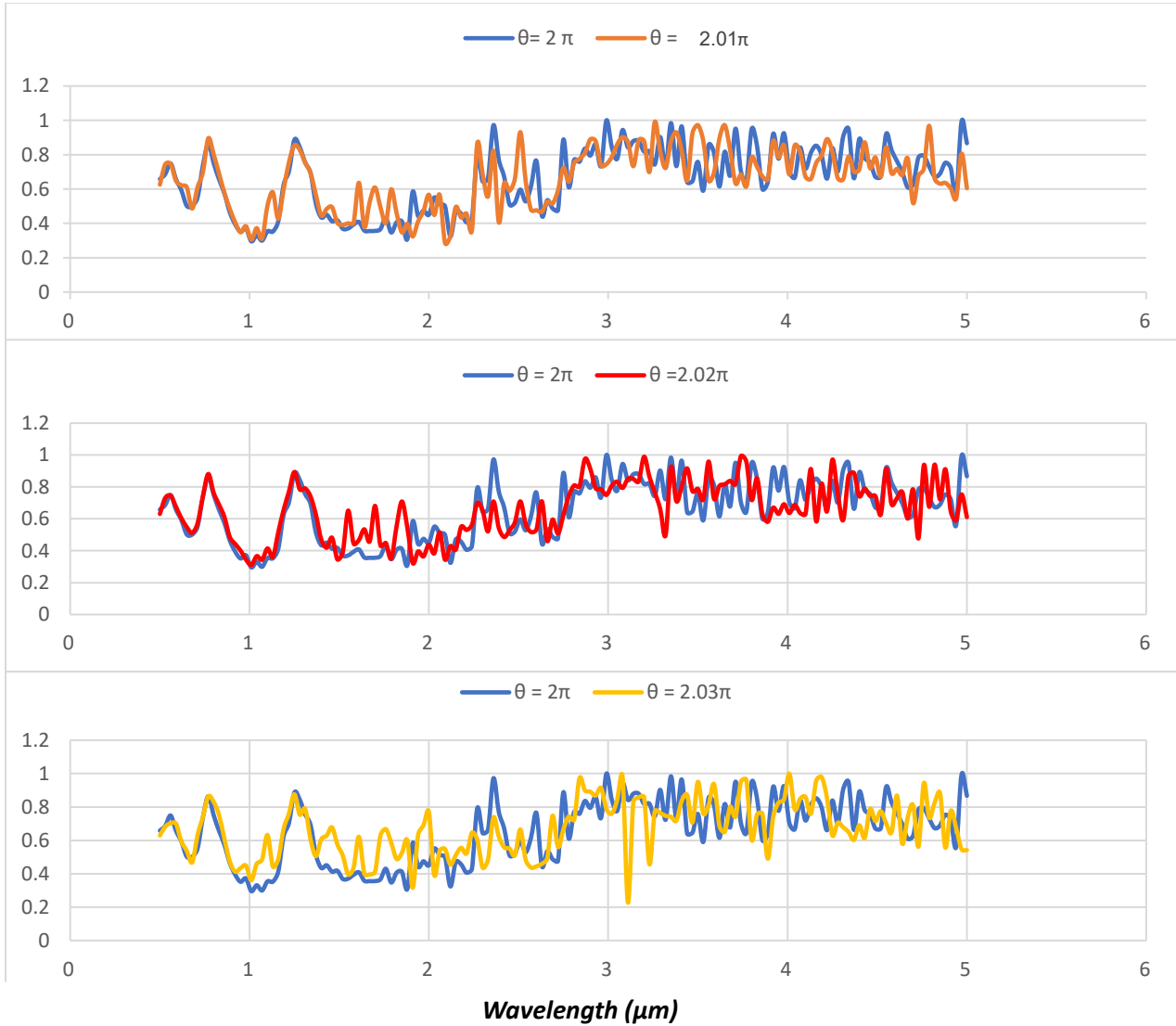

**Fig 11.** The effects of increasing the θ parameter of the Archimedean spiral (r = a + b*θ) on the absorption spectrum of the metamaterial absorber

In Figure 11, we investigate the effects of increasing the θ parameter of the Archimedean spiral, where r = a + b*θ, on the absorption spectrum of the metamaterial absorber. By systematically increasing the θ parameter, we aim to understand how this parameter influences the absorption characteristics of the absorber. We observe the effects of increasing the θ (theta) parameter in the absorbent structure on the absorption spectrum of the metamaterial absorber. Specifically, we compare three different θ values, namely 2.01π, 2.02π, and 2.03π, with the initial state where θ is set to 2π. When θ increases from 2π to 2.01π, we observe absorption percentages of 60%, 74%, 75%, 81%, and 73% for wavelengths of 2.51 µm, 3.26 µm, 3.5 µm, 3.65 µm, and 4.79 µm, respectively. As we increase θ, these absorption percentages rise to 93%, 99%, 97%, 97%, and 96%, respectively. Increasing θ from 2π to 2.02π results in absorption percentages of 83%, 81%, 85%, 64%, 83%, 79%, and 67% for wavelengths of 2.87 μm, 3.2 μm, 3.56 μm, 3.77 μm, 4.25 μm, 4.76 μm, and 4.82 μm, respectively. With the increased θ, these percentages increase to 97%, 98%, 95%, 96%, 95%, 97%, 93%, and 94%, respectively. Finally, when θ increases from 2π to 2.03π, we observe absorption percentages of 75%, 64%, 70%, 79%, and 79% for wavelengths of 2.84 μm, 3.77 μm, 4.01 μm, 4.19 μm, and 4.76 μm, respectively. As θ increases, these percentages rise to 97%, 96%, 99%, 97%, and 94%, respectively.

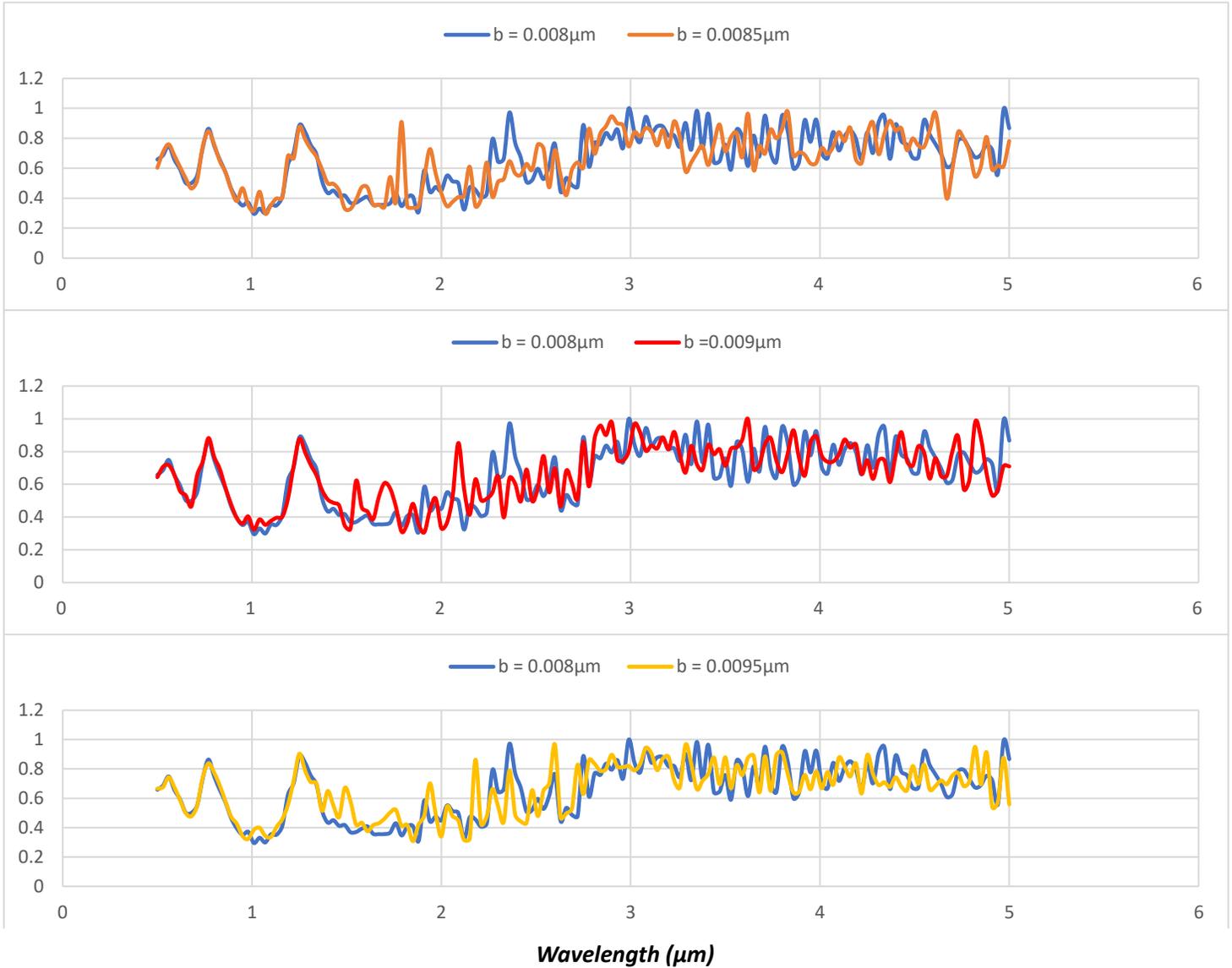

**Fig 12.** The effects of increasing the b parameter of the Archimedean spiral (r = a + b*θ) on the absorption spectrum of the metamaterial absorber

In Figure 12, we investigate the effects of increasing the b parameter of the Archimedean spiral, where r = a + b*θ, on the absorption spectrum of the metamaterial absorber. Specifically, we compare three different b values, namely 0.0085 μm, 0.009 μm, and 0.0095 μm, with the initial state where b is set to 0.008 μm. When b increases from 0.008 μm to 0.0085 μm, we observe absorption percentages of 35%, 87%, 57% and 73% for wavelengths of 1.79 μm, 2.9 μm, 3.62 μm and 4.61 μm, respectively. As we increase b, these absorption percentages rise to 90%, 90%, 96%, and 97%, respectively. Increasing b from 0.008 μm to 0.009 μm results in absorption percentages of 74%, 83% and 67% for wavelengths of 2.9 μm, 3.62 μm and 4.82 μm, respectively. With the increased b, these percentages increase to 98%, 99.7%, 95% and 98%, respectively. Finally, when b increases from 0.008 μm to 0.0095 μm, we observe absorption percentages of 75%, 42%, 70% and 69% for wavelengths of 2.18 μm, 2.6 μm, 3.29 μm and 4.82 μm, respectively. As θ increases, these percentages rise to 86%, 96%, 96% and 95%, respectively.

And finally, we investigate the size variation of parameter "a" in the Archimedean structure of our metamaterial absorber to assess its impact on the absorption peak of the absorption spectrum in our designed metamaterial.

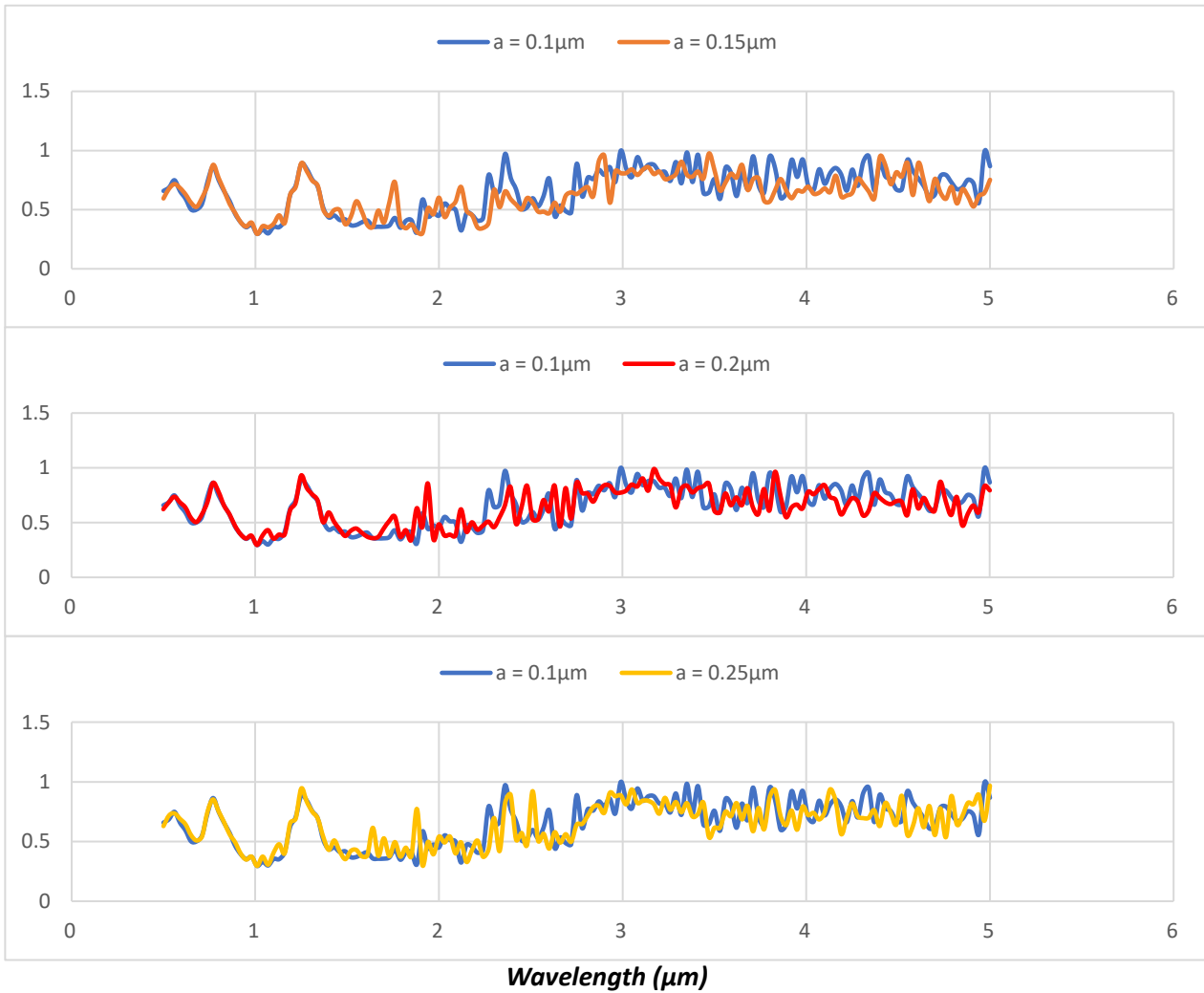

*Wavelength (μm)*

**Fig 12.** The effects of increasing the a parameter of the Archimedean spiral (r = a + b*θ) on the absorption spectrum of the metamaterial absorber

In Figure 13, we investigate the effects of increasing the a parameter of the Archimedean spiral, where r = a + b*θ, on the absorption spectrum of the metamaterial absorber. Specifically, we compare three different a values, namely 0.15 μm, 0.2 μm, and 0.25 μm, with the initial state where a is set to 0.1 μm. When a increases from 0.1 μm to 0.15 μm, we observe absorption percentages of 90% and 54% for wavelengths of 2.87 μm and 3.47 μm, respectively. As we increase b, these absorption percentages rise to 95% and 97%, respectively. Increasing a from 0.1 μm to 0.2 μm results in absorption percentages of 48%, 53% and 84% for wavelengths of 1.94 μm, 2.48 μm and 3.17 μm, respectively. With the increased b, these percentages increase to 85%, 84% and 98%, respectively. Finally, when b increases from 0.1 μm to 0.25 μm , we observe absorption percentages of 53% and 84% for wavelengths of 2.51 and 4.13 μm, respectively. As a increases, these percentages rise to 92% and 93%, respectively.

## Conclusion

In this paper, we have presented a tunable metamaterial absorber that effectively controlled the absorption spectrum by manipulating the conductivity of vanadium oxide. The absorber incorporated an Archimedean structure, which yielded multiple absorption peaks, demonstrating enhanced absorption performance within the wavelength range of 0.5 to 5 micrometers. Through comprehensive analysis, we examined the influence of various parameters, including changes in the angle of incidence and azimuthal variations, and modifications to the Archimedean structure's parameters (a, b, θ, and thickness S), on the absorption spectrum. Our findings indicate that the proposed with Archimedean structure achieved higher absorption percentages and exhibited absorption peaks that approached complete absorption. This highlights the potential of the Archimedean spiral structure as a promising design element in constructing metamaterials for specific applications. The versatility and tunability demonstrated in our study open up new possibilities for its utilization in various fields.

## Acknowledgments

The authors would like to express their gratitude to Mazandaran University for providing the computational resources required for the simulations conducted in this article.

## Data Availability

Data underlying the results presented in this paper are not publicly available at this time but may be obtained from the authors upon reasonable request.

## Compliance with Ethical Standards

## Competing Interests

The authors declare that they have no competing interests.

## Funding information

This study hasn't any financial support.

## Authors' Contributions

All authors have same contribution in the analytical and numerical calculations and read and approved the final manuscript.